\documentclass[12pt,preprint]{aastex}

\def\arcsecpoint{$''\!.$}
\def\deg{$^{\rm o}$}

\def\gtsim{\raisebox{-.5ex}{$\;\stackrel{>}{\sim}\;$}}

\shortauthors{Crenshaw \& Kraemer}
\shorttitle{The NLR and UV Absorbers}

\begin{document}

\title{The Connection between the Narrow-Line Region and the UV Absorbers in 
Seyfert Galaxies\altaffilmark{1}}

\author{D.M. Crenshaw}
\affil{Department of Physics and Astronomy, Georgia State University,
Atlanta, GA 30303; crenshaw@chara.gsu.edu}

\author{S. B. Kraemer}
\affil{Catholic University of America and Laboratory for Astronomy and
Solar Physics, NASA's Goddard Space Flight Center, Code 681,
Greenbelt, MD  20771; stiskraemer@yancey.gsfc.nasa.gov}

\altaffiltext{1}{Based on observations made with the NASA/ESA Hubble Space 
Telescope, obtained at the Space Telescope Science Institute, which is 
operated by the Association of Universities for Research in Astronomy, Inc., 
under NASA contract NAS 5-26555.}

\begin{abstract}

We present evidence that the outflowing UV absorbers in Seyfert 1 galaxies
arise primarily in their inner narrow (emission) line regions (NLRs), based on
similarities in their locations, kinematics, and physical conditions. 1) {\it
Hubble Space Telescope} observations show that nearly all Seyfert galaxies have
bright, central knots of [O~III] emission in their NLRs with radii of tens of
parsecs. These sizes are consistent with most previous estimates of the
distances of
UV (and X-ray) absorbers from their central continuum sources, and a
recently-obtained reliable distance of $\sim$25 pc for a UV absorber in the
Seyfert 1 galaxy NGC 3783. 2) The nuclear emission-line knots in a sample of 10
Seyfert galaxies have velocity widths of 300
-- 1100 km s$^{-1}$ (half-width at zero intensity), similar to the radial
velocities of most UV absorbers. The highest radial velocity for a Seyfert UV
absorber to date is only $-$2100 km s$^{-1}$, which is much lower than typical
broad-line region (BLR) velocities. There is also mounting evidence that the
NLR clouds are outflowing from the nucleus, like the UV absorbers. 3) If our
hypothesis is correct, then the NLR should have a component with a high global
covering factor ($C_g$) of the continuum source and BLR, to match that found
from previous surveys of UV absorbers ($C_g$ = 0.5 -- 1.0). Using STIS spectra
of NGC~4151, obtained when the continuum and BLR fluxes were low, we find
evidence for optically thin gas in its nuclear emission-line knot. We are able
to match the line ratios from this gas with photoionization models
that include a component with $C_g$ $\approx$ 1 and an ionization
parameter and hydrogen column density that are typical of UV absorbers.

\end{abstract}

\keywords{galaxies: Seyfert -- ultraviolet:galaxies}
~~~~~

\section{Introduction}

Mass outflows of ionized gas from active galactic nuclei (AGN) have been
detected in the majority of Seyfert 1 galaxies observed in the UV
(Crenshaw et al. 1999; Kriss 2002) and X-rays (Reynolds 1997, George et al.
1998). The outflows are revealed through absorption lines that are blueshifted
with respect to the systemic velocities of the host galaxies. Observations at
high spectral resolution ($\lambda$/$\Delta\lambda$ $\gtsim$ 10,000) in the UV
show that the absorption tends to split into multiple kinematic components
with widths in the range 20 -- 500 km s$^{-1}$. Detailed studies of
the ``intrinsic'' absorbers indicate large global covering factors ($C_g$
$=$ 0.5 -- 1.0) of the nucleus and mass-loss rates compable to the inferred mass
accretion rates ($\sim$0.01 M$_{\odot}$ yr$^{-1}$) (Crenshaw et al.
2003a). The origin of the intrinsic absorbers (e.g., accretion disk, torus) and
the means by which they are accelerated outward (e.g., thermal winds,
radiatively-driven flows, hydromagnetic flows) are currently unknown.

The large global covering factors of the UV absorbers suggest that they should
contribute to at least some of the UV and optical emission that we see. Thus, it
is only natural to ask if they are associated in some way with either the narrow
(emission) line region (NLR) or the broad (emission) line region (BLR). An
important clue is that the highest radial velocity observed in any Seyfert 1
galaxy is only$-$2100 km s$^{-1}$ (Crenshaw et al. 1999), which is comparable
to the maximum velocity (half-width at zero intensity) of the NLR gas and much
less than that of the BLR gas, which extends out to $\gtsim$10000 km s$^{-1}$.
Furthermore, the available evidence indicates that UV absorbers have transverse
velocities comparable to their radial velocities (Crenshaw et al.
2003a, 2004a). Thus, the kinematics of the absorbers suggest that they are more
closely related to the NLR than the BLR.

Another clue to the origin of the absorbers and their connection to the
emission-line regions is their radial location with respect to the central
supermassive black hole (SMBH), but this is a difficult parameter to obtain.
Early studies of the ultraviolet spectrum of NGC~4151 with the {\it
International Ultraviolet Explorer} found that the UV absorbers must lie
outside of the broad (emission) line region (BLR), since both continuum and
broad-line emission were absorbed (Bromage et al. 1985). This was later found
to be the case as well for a sample of Seyfert 1 galaxies observed by the {\it
Hubble Space Telescope} ({\it HST}) (Crenshaw et al. 1999). Various studies have
placed upper or lower limits on the distances of outflowing UV or X-ray
absorbers from their nuclei, and these fall in the range of tenths to tens of
parsecs\footnote{By contrast, intrinsic absorption lines that are within a few
hundred km s$^{-1}$ of the systemic velocity can often, but not always, be
attributed to gas at much larger distances in the host galaxy's disk or halo
(Weymann et al. 1997; Crenshaw \& Kraemer 2001).} (Kriss et al. 1997; Espey et
al. 1998; Netzer et al. 2002; Kraemer et al. 2002; Gabel et al. 2003a; Netzer et
al. 2003).

Recently, we determined a reliable distance for an outflowing UV absorber (Gabel
et al. 2004a), which is Component ``1'' in NGC 3783, at a radial velocity of
$-$1350 km s$^{-1}$ with respect to the systemic velocity of the host galaxy.
From the relative strengths of the metastable
C~III* $\lambda$1175 multiplet lines in absorption, we determined that the
hydrogen number density of the absorber was $n_H$ $\approx$ 4 $\times$
10$^{4}$ cm$^{-3}$. Combined with the ionization parameter and ionizing
luminosity, this yields a distance of $\sim$25 pc from the central continuum
source (Gabel et al. 2004a). This absorber is therefore in the inner NLR, which
typically extends out to several hundred parsecs from the nucleus
(Schmitt et al. 2003a, b).

Given their locations and velocities, are the UV absorbers just NLR clouds seen
in absorption? Global estimates of the covering factor of the NLR in Seyfert 1
galaxies yield values of $C_g$ $\approx$ 0.02 (Netzer \& Laor 1993).
Furthermore, NLR clouds tend to have a lower ionization parameter ($U$), defined
as the ratio of ionizing photons to hydrogen atoms at the ionized face, and
higher hydrogen column density ($N_H$) than UV absorbers, on average
(Crenshaw et al. 2003a).  However, previous studies have already shown that a
typical NLR is composed of a multi-phase medium; clouds with different physical
conditions ($U$, $N_H$, $n_H$) coexist at the same radial locations
(Kraemer \& Crenshaw 2000, Kraemer et al. 2000b). One possibility is that
UV absorbers are associated with a high-ionization, diffuse component of the
NLR, rather than the entire NLR. This idea is similar to that of Cecil et al.
(2002), who suggest that the high-velocity (extending up to $-$3200 km
s$^{-1}$) emission-line clouds in the NLR of NGC 1068 would resemble the
``associated absorbers'' in quasars if seen against the continuum source.
However, these high-velocity clouds have a covering factor of $C_g$ $\leq$
0.02. In this paper, we investigate the possibility that there is a
high-ionization component of the NLR with a covering factor that is
sufficiently large to account for most of the intrinsic UV absorption in
Seyfert galaxies.

{\it HST} images of the NLR show that nearly all Seyfert 1 and 2
galaxies have a bright central knot of [O~III] emission (Schmitt et al. 2003a),
and Space Telescope Imaging Spectrograph (STIS) slitless spectra of these
nuclear knots show they have high velocity
dispersions (Ruiz et al. 2005). Their high surface brightnesses, proximity to
their nuclei, and high velocity dispersions indicate
they are good candidates for UV absorbers seen in emission. In this paper, we
explore a number of their properties based on the STIS spectra of Ruiz et al.
(2005). To demonstrate that the nuclear emission-line knots can be responsible
for the UV absorption, it is necessary to show that their spectra show evidence
for a high-ionization, high covering factor component. We present a
STIS UV/optical spectrum of the nuclear [O~III] knot in NGC~4151 that is
ideal for this purpose (\S4). First, however, we provide new measurements that
explore the kinematic and positional similarities between the UV absorbers (\S2)
and the NLR
(\S3).

\section{Kinematics of the UV Absorbers}

Observations at high spectral resolution ($\lambda$/$\Delta\lambda$ $\geq$
10,000) are needed to resolve the kinematic components of intrinsic absorption
in Seyfert galaxies. This requirement is met in the UV (1200 -- 3200 \AA) by
observations with the high-resolution modes of {\it HST}'s STIS or the
first-generation Goddard
High-Resolution Spectrograph (GHRS), and in the far-UV (900 -- 1200 \AA) with
the {\it Far Ultraviolet Spectroscopic Explorer} ({\it FUSE}). We have compiled
the available measurements of the radial-velocity centroid ($v_r$), relative to
the host galaxy, and full-width at half-maximum (FWHM) of the absorption depth
for individual absorption components observed by these instruments. The sample
contains 61 absorption components in 10 Seyfert 1
galaxies: NGC 3516 (Crenshaw et al. 1999; Kraemer et al. 2002), NGC 3783
(Kraemer et al. 2001a; Gabel et al. 2003), NGC~5548 (Mathur et al. 1999;
Crenshaw et al. 2003), Mrk~509 (Kriss et al. 2000; Kraemer et al. 2003),
NGC~4051 (Collinge et al.
2001), NGC 4151 (Weymann et al. 1997; Kraemer et al. 2001b), Mrk 279 (Scott et
al. 2004; J.R. Gabel, 2004, private communication), NGC~7469 (Kriss et al.
2003), Akn 564 (Crenshaw et al. 2002), and NGC 4395 (Crenshaw et al. 2004b).

Figure 1 plots the FWHM vs. $v_r$ for each intrinsic absorption component in our
sample. Although the sample is rather sparse, there is no clear trend;
components at both high and low $|v_r|$ show a large range in FWHM. The
tendency for most low (within $\pm$ 250 km s$^{-1}$) $v_r$ components to have
low ($<$~100 km s$^{-1}$) FWHM reflects the finding that a large fraction arise
in the Seyfert host galaxies, as has been demonstrated for specific components
in NGC 4151 (Weymann et al. 1997), Akn 564 (Crenshaw et al. 2002), and NGC 4395
(Crenshaw et al. 2004b). Furthermore, a galactic origin may very well explain
all of the
cases of positive $v_r$. However, some of these low $v_r$ components are likely
intrinsic to the AGN; for example, a component in NGC~3516 at $v_r$ $=$ 30 km
s$^{-1}$ has shown strong variability in its column density, indicating that it
arises close to the nucleus (Kraemer et al. 2002).
\footnote{There are several other aspects of Figure 1 that deserve comment. 
The FWHMs have not been corrected for instrumental resolution ($\sim$15 km
s$^{-1}$ for {\it FUSE}, $\sim$8 km s$^{-1}$ for STIS), so that lines with the
lowest FWHM (20 -- 30 km s$^{-1}$) are just barely resolved. Some of the
absorption lines are likely saturated, so the FWHM may overestimate their
intrinsic velocity spread.
The four points with low FWHM at $\sim$ $-$2100 km s$^{-1}$ are subcomponents of
an absorption feature in NGC~7469 (Kriss et al. 2003). The point with very
large FWHM (940 km s$^{-1}$) represents the transient absorption component
D$'$ in NGC 4151 (Kraemer et al. 2001a).}

For our purposes, the importance of Figure 1 is that it shows the radial
velocities of intrinsic UV absorbers are $|v_r| \leq$ 2100 km s$^{-1}$ in
Seyfert galaxies, and the majority of components (51 out of 61) have $|v_r| <$
1000
km s$^{-1}$. In a survey of low-resolution UV spectra of Seyfert galaxies, we
found a similar result: the highest (most negative) radial velocity was
$-$2150 km s$^{-1}$, for I~Zw~1 (Crenshaw et al. 1999). Transverse velocities of
the absorbers
are more difficult to obtain, but a few lower limits on the order of $\sim$1000
km s$^{-1}$ have been determined by detecting the motion of absorbing clouds
across the BLR (Crenshaw et al. 2003a, and references therein). The only
reasonably
accurate transverse velocity to date is for ``Component 1'' in NGC~3783, which
lies in the range $v_T$ $=$ 540 -- 1430 km s$^{-1}$ (Crenshaw, Kraemer, \& Gabel
2004a). Many other absorption components have been monitored on a yearly
basis and have not shown evidence for transverse motion, so the transverse
velocities could be much lower for these components. Although
more work needs to be done, the available evidence indicates that the space
velocities of UV absorbers in Seyfert galaxies are much less than those of BLR
clouds, and very similar to the velocities in the NLR.

\section{Kinematics and Sizes of the Nuclear Emission-Line Knots}

We require the high spatial resolution of {\it HST} to isolate the nuclear
emission-line knots and measure their sizes, and the spectral resolution
of STIS to accurately measure their kinematics. STIS medium-dispersion
spectra of the bright [O~III] $\lambda$5007 emission line are ideal for this
purpose. We use the STIS slitless spectra of NGC 4151 (Hutchings et al. 1998;
Kaiser et al. 2000) and nine other Seyfert galaxies (Ruiz et al. 2005; see this
paper for details on the observations and data reduction),
obtained with the G430M grating and an open aperture. The spectra have a
spatial resolution of 0\arcsecpoint1 (full-width at half-maximum) (FWHM) and a
velocity resolution of 33 km s$^{-1}$ (FWHM) in the dispersion direction.
Ruiz et al. (2005) found that nine of the ten Seyfert galaxies in their sample
show bright, compact (FWHM $<$ 0\arcsecpoint5) [O~III] knots at their optical
continuum peaks with apparently high velocity dispersions. The only exception
is Mrk~3, which shows {\it several} bright knots with high velocity dispersions
within $\sim$1$''$ of its optical nucleus; we do not use Mrk~3 for this study.

For the 10 nuclear emission-line knots, we measured the half-width at
half-maximum (HWHM) of the [O~III] $\lambda$5007 emission in the dispersion
(radial velocity) and cross-dispersion (spatial)
directions. We also measured the half-width at zero intensity (HWZI) in the
dispersion direction. We use these parameters rather than the
full widths to compare with the UV absorber kinematics, because they are more
appropriate for simple geometries. For an expanding shell, for example, the
HWHM of the intensity profile in the spatial direction can be compared to the
average absorber distance, and the HWZI in the spectral direction can be
compared to the maximum absorber radial velocities. However, we note that direct
comparisons can only be made by assuming very specific geometric and kinematic
models.

Table 1 shows measurements of the nuclear knots in our sample, which
includes two Seyfert 1, one Seyfert 1.9, and seven Seyfert 2 galaxies at
redshifts $<$ 0.03. Three of the nuclear knots are only marginally resolved,
and we list their HWHM sizes (in arcsecs and parsecs) as upper limits. The
HWHM's range from 7 to 67 pc, similar to estimates for the distances of UV
absorbers from their continuum sources, as described in \S1. The HWZI's range
from 310 to 1060 km s$^{-1}$, similar to the range of maximum absorber
velocities found in most Seyfert galaxies, as described in \S2.

\section{Physical Conditions in the Nuclear Emission-Line Knot of NGC~4151}

\subsection{Observations and Emission-Line Ratios}

To determine the physical conditions in a nuclear emission-line knot, we would
like to have
extensive UV and optical coverage to get as many emission-line diagnostics as
possible. We also require the high spatial resolution of {\it HST} to isolate
the knot. Of the sources in Table 1, only NGC~4151 has low-dispersion STIS
spectra covering a large wavelength range (1150 -- 10,270 \AA). We searched the
Multimission Archives at the Space Telescope Science Institute (MAST) and found
two sets of low-dispersion spectra of the nucleus: one
obtained on 1998 January 8/February 10 (published in Nelson et al. 2000) and
one obtained on 2000 May 24/28 (previously unpublished). The latter set was
obtained when the UV and BLR fluxes were in a much lower state (continuum flux
at 1465 \AA\ $=$ 2.0 ($\pm$0.3) $\times$ 10$^{-14}$ ergs s$^{-1}$ cm$^{-2}$
\AA$^{-1}$) compared to that of the 1998 observations (24.1 ($\pm$0.5)
$\times$ 10$^{-14}$ ergs s$^{-1}$ cm$^{-2}$ \AA$^{-1}$), and is
therefore ideal for deconvolving emission lines with both broad and narrow
components. The only drawback in using NGC~4151 is that it has strong intrinsic
absorption lines that contaminate the narrow emission-line spectrum in the UV.
In particular, absorption due to the host galaxy and/or halo (components F and
F$'$ in Weymann et al. 1997) severely absorb the narrow
emission components of the UV resonance lines of L$\alpha$, N~V, O~I, C~II,
Si~IV, C~IV, and Mg~II, to the extent that their NLR contributions are not
measurable even in STIS echelle spectra (see Kraemer et al. 2001b). However,
when NGC~4151 is in a high state, the unusual (for Seyferts) Balmer absorption
lines nearly
disappear (Hutchings et al. 2002). Thus, we use the high-state slitless
spectra of H$\beta$ and H$\alpha$ obtained on 1997 July 15 (previously
published in Kaiser et al. 2000; Hutchings et al. 2002) to deconvolve their
narrow and broad emission components.

In Table 2, we list the STIS observations that we retrieved from the MAST
for this analysis. The first four observations are the low-dispersion,
long-slit spectra, which have a resolving power of $\lambda$/$\Delta\lambda$
$\approx$ 1000, and the last two are the medium-dispersion, slitless spectra
of H$\beta$ (plus [O~III] $\lambda\lambda$ 4959, 5007) and H$\alpha$ (plus
[N~II] $\lambda\lambda$6548, 6584), with $\lambda$/$\Delta\lambda$ $\approx$
9000. We reduced the STIS low- and medium-dispersion spectra using the IDL
software developed at NASA's Goddard Space Flight Center for the Instrument
Definition Team. We identified and removed cosmic-ray hits using multiple
images and removed hot or warm pixels by interpolation in the dispersion
direction. We used wavelength calibration exposures obtained after each science
observation to correct the wavelength scale for zero-point shifts. For the
G750L spectra, we used a contemporaneous continuum-lamp exposure in the
flat-fielding process to remove the fringes at the long-wavelength end. We
extracted the spectra from the flux-calibrated CCD images using a bin height of
11 pixels (0\arcsecpoint55) in the cross-dispersion direction, which
corresponds to the full-width at zero-intensity (FWZI) of the [O~III] nuclear
knot from the G430M slitless spectrum. For the flux-calibrated MAMA images, we
used the same extraction height of 0\arcsecpoint55, which corresponds to
22 pixels on the MAMA detector. We combined the low-dispersion spectra in their
regions of overlap to obtain a single spectrum from 1150 -- 10,270 \AA.

In Figure 2, we show the full low-dispersion spectrum of the nuclear
emission-line knot in NGC~4151. Narrow emission lines covering a broad range in
ionization are present, from [O~I] to [Fe~XI]; the ionization potential needed
to create the latter is IP$_c$ $=$ 262 eV. We see no evidence for even higher
ionization lines, like the [S~XII] $\lambda$7611 line (IP$_c$ $=$ 505 eV) seen
in the STIS spectrum of the hot spot in NGC~1068 (Kraemer \& Crenshaw 2000a).
The broad components of the permitted lines are clearly evident but still
unusually weak for NGC~4151. Although UV resonance lines are
detected in emission, their fluxes are severely affected by the intrinsic
absorption lines, as discussed previously.

We measured the fluxes of isolated emission lines by direct integration above
local continuum fits. For narrow lines that
were blended with other broad or narrow emission, we used the [O~III]
$\lambda$5007 profile as a template to deconvolve the blends (see Crenshaw \&
Peterson 1986). We used the medium-dispersion spectra to deconvolve the blend
of narrow and broad H$\beta$ and the blend of narrow and broad H$\alpha$ and
the [N II] lines. As shown in Figure 3, the narrow component of H$\beta$, which
we use as our standard for line ratios, was easily separated from the broad
component in this medium-dispersion spectrum.
Lines that we deblended in the low-dispersion spectra include the broad and
narrow components of the H and He permitted lines and the semi-forbidden lines
in the UV, as well as the narrow [S~II] $\lambda\lambda$6716, 6731 lines and
H$\gamma$ plus [O~III] $\lambda$4363 lines. We determined the reddening of the
the narrow emission lines from the He~II $\lambda$1640/$\lambda$4686 ratio, the
Galactic reddening curve of Savage \& Mathis (1979), and an intrinsic He~II
ratio of 7.2. We determined uncertainties in the dereddened ratios from the sum
in quadrature of the errors from three sources: photon noise, different
reasonable continuum placements, and reddening.

Table 3 gives the observed and dereddened narrow-line ratios, relative to
H$\beta$, as well as the uncertainties in these ratios. The reddening
determined from the He~II ratio is only E(B$-$V) $=$ 0.02 $\pm$ 0.04. The
reddening correction has a very small effect on our line ratios, as shown in
Table 3, and uncertainties in the correction have a negligible effect on our
results. The reddening that we have determined can
be attributed entirely to reddening in our Galaxy, which is E(B$-$V) $=$ 0.03
(Schlegel et al. 1998). Thus, there is no evidence for dust in the inner NLR of
NGC 4151, although at other locations in the NLR, the observed reddening of the
emission lines ranges from E(B$-$V) $=$ 0.01 to 0.36 (Kraemer et al. 2001b). 

\subsection{Photoionization Models}

To explore the physical conditions and covering factor of the nuclear knot in
NGC~4151, we generated photoionization models using the code Cloudy (Ferland et
al. 1998). We modeled the emission-line gas as single-zoned slabs, directly
irradiated by the central source. We used the same spectral energy distribution
(SED) of the ionizing continuum radiation as we used previously for NGC~4151
(Kraemer et al. 2001b), which consist of broken power-laws of the form $F_{\nu}
\propto \nu^{\alpha}$, where the spectral index $\alpha = -1.0$ for h$\nu$ $<$
13.6 eV, $\alpha = -1.4$ over the range 13.6 eV $\leq$ h$\nu$ $<$ 1 keV, and
$\alpha = -0.5$ above 1 keV, and the luminosity in ionizing photons is 2
x 10$^{53}$ s$^{-1}$. We assumed roughly solar elemental abundances (e.g.
Grevesse \& Anders 1989) and that the gas was free of cosmic dust. The models
are parameterized in terms of the ionization parameter ($U$), the hydrogen
number density ($n_H$), and the hydrogen column density ($N_H$). 

The emission-lines in Table 3 span a huge range in ionization and critical
density, and it is clear that multiple model components are required. Since
the nuclear knot is resolved, we used the radius inside of which one-half of the
knot's [O~III] emission is contained as a starting point for the distances of
the components. This radius is 0\arcsecpoint15, which corresponds to a
projected distance of 9.5 pc from the central continuum source.

The dereddened He~II $\lambda$4686/H$\beta$ ratio in Table 3 is relatively high
for photoionized gas, which suggests either 1) a much harder continuum SED than
we have used, or 2) a strong contribution from a high-ionization, matter-bound
(optically thin to hydrogen-ionizing radiation) component. There is no evidence
for a much harder SED and, in fact, we have used the above SED to successfully
model the emission-line ratios from other parts of the NLR in NGC~4151 (Kraemer
et al. 2000). Furthermore, any SED that possesses a realistic EUV bump would
boost both the He II and [Ne V] emission, but what is required is a high
[Ne~V]/He~II ratio (Table 3). This can only be achieved with an optically thin 
component that has a large Ne$^{+4}$ zone prior to the point where the gas
becomes optically thick to the He~II ionizing radiation (Kraemer \&
Crenshaw 2000; Kraemer et al. 2000).

We have therefore adopted a matter-bound component (``UVABS'') with
parameters typical of most UV absorbers (Crenshaw et al. 2003): $U = 0.03$ and
$N_H =$ 10$^{19.5}$ cm$^{-2}$ (with the above parameters, $n_H$ is fixed to
10$^{4.3}$ cm$^{-3}$). The predicted ionic column densities listed in the
footnotes  to Table 3 would result in absorption lines that are typical of most
UV absorbers:
strong-to-saturated Ly$\alpha$, C~IV $\lambda\lambda$ 1548.2, 1550.8,
N~V $\lambda\lambda$ 1238.8, 1242.8, and O~VI $\lambda\lambda$1031.9, 1036.3,
and weak or undetectable Si~IV $\lambda\lambda$1393.8, 1402.8.

In order to fit the observed emission-line ratios, we required two additional
components: a moderate density, radiation-bound component (``OIII'') to
reproduce the strong [O~III] $\lambda$5007 emission, and a higher density,
radiation-bound component (``DENSE'') to account for 
the strong [O~III] $\lambda$4363 line (see Table 3). We
were unable to find a satisfactory single-component solution, since, for
example, the [Ne~IV] $\lambda$ 2423 line is quenched at intermediate densities.
The OIII component was kept at a distance of 9.5 pc, consistent with
measurements of the [O~III] $\lambda$5007 emission. We kept the same ionization
parameter ($U
=$ 0.01) for
the DENSE component, but increased its density, which effectively moves it to a
distance of 2.0 pc from the continuum source.
To produce a composite model spectrum, we adjusted the weight of each component
according to the fraction of H$\beta$ it contributes until we obtained a good
match with the dereddened line ratios. We give the component and composite line
ratios in Table 3, along with the final model parameters, the weight of each
component, and the flux of H$\beta$ at the ionized face of each component
($f_{H\beta}$).

Nearly all of the dereddened line ratios are fit well by the composite model.
The low critical density lines (see
Osterbrock 1989) from low ionization species such as [O~II] $\lambda$3227 and
[S~II] $\lambda\lambda$ 6716, 6731 are underpredicted, which could be rectified
by introducing an additional component outside the emission-line bicone that is
irradiated by a heavily absorbed continuum, similar to the model
described for the hot spot in NGC 1068 (Kraemer \& Crenshaw 2000a). Also,
the model significantly underpredicts the strengths of the [Fe~VII] and 
[Fe~XI] lines, but this is a common problem in NLR models (Oliva 1997) that may
be due to problems with the atomic data (S.B. Kraemer \& G.J. Ferland, 2004, in
preparation).

Table 3 shows that the inclusion of the UVABS component allows us to match the
high He~II/H$\beta$ ratio. UVABS also provides a significantly better fit to
the other He~II lines and the high-ionization N~IV], [Ne~IV], and [Ne~V] lines
(note that He~II $\lambda$3204 is in a noisy region of the spectrum and has a
large uncertainty)
\footnote{To explore the possible contribution to He~II from a collisionally
ionized 
plasma, we assumed the gas is in rough thermal equilibrium with our emission
line components, and derived a density n(H) = 10$^{2.5}$ cm$^{-3}$. This
permits a column density of $\sim$10$^{22}$ cm$^{-2}$ within the 
central knot. We then generated a model of collisionally ionized plasma with 
these initial conditions using Cloudy90. The predicted emitted He~II
$\lambda$4686 flux was 4.27 $\times$ 10$^{-3}$ ergs cm$^{-2}$ s$^{-1}$, which
is 0.08 the contribution from UVABS, or 0.028 of the total observed He~II.
Hence, this component, which, of course would contribute less if the
temperature were higher, cannot contribute significantly to the observed He II
emission.}.
Even with the inclusion of UVABS, our model underpredicts the [Ne~V]/H$\beta$
ratio somewhat. This cannot be corrected with a harder SED because it would
lead to an overprediction of He~II/H$\beta$, as discussed earlier in this
section. One plausible explanation is that the ionizing continuum has been
modified by optically thin gas closer to the nucleus, which has absorbed some
of the radiation near the He~II Lyman limit but has no effect at energies
$\geq$ 97 eV (the ionization potential of Ne$^{+3}$). The result would be that
the [Ne~V] would be enhanced with respect to He~II, as demonstrated in Kraemer
et al. (2000). 

Based on our model predictions, we can constrain the covering factors of the
individual components. The extinction-corrected
H$\beta$ luminosity of the central knot is 6.4 x 10$^{39}$ erg
s$^{-1}$. Based on the relative contribution to H$\beta$ from each model and
the emitted flux of H$\beta$ at the ionized face (Table 3), we derive the
following covering fractions: for  UVABS, $C_g$ $=$ 1.1; for OIII, $C_g$
$=$ 0.02; and for DENSE, $C_{g}$ $=$ 0.03.

The global covering factor of the UVABS component is close to one,
which, given our assumptions, is consistent with the derived value of
$C_g$ $=$ 0.5 -- 1.0 for intrinsic UV absorption in Seyfert galaxies (Crenshaw
et al. 1999). Increasing the column density $N_H$ of UVABS would decrease
$C_g$, but this would decrease the He~II/H$\beta$ ratio significantly below the
observed value
\footnote{One component in NGC~4151 (``D$+$E'' in Kraemer et al. 2001b) shows
evidence for partial covering in the line-of sight, which indicates partial
global covering. However, this component is much closer to the source and has a
much higher column density than UVABS; there are other UV absorption components
in NGC~4151 that may fully cover the continuum source.}.
We conclude that there is strong evidence for a
high-ionization, matter-bound component in the nuclear knot of NGC~4151 with a
high global covering factor.

\section{Discussion}

We have shown that the bright central knots of emission in the NLRs of most
Seyfert galaxies are likely sources of intrinsic absorption lines that we
detect in their UV spectra, particularly absorbers that are outflowing with
radial velocities
$v_r < -300$ km s$^{-1}$. Our evidence is based on the similarities in
the kinematics and locations of the NLR knots and the UV absorbers.
More work must be done to test these connections. In particular, more
information is needed on the distances of the UV absorbers from their central
continuum sources, based on monitoring campaigns and/or detection of metastable
lines (see Crenshaw et al. 2003a). We note that intrinsic absorption lines that
are within a few hundred km s$^{-1}$ of the systemic velocity can arise from
either the AGN or host galaxy; the percentage from each is not known and
deserves further study.

To establish a more direct link between the emission and absorption, we modeled
the narrow emission lines from the resolved central knot in NGC~4151.
We found that the line ratios indicate a high-ionization, matter-bound component
in the nuclear emission-line knot. Our characterization of this component in
the form of ``UVABS'' yields a high global covering factor, which matches
that found for intrinsic UV absorption in Seyfert 1 galaxies ($C_g \approx
0.5 - 1$). Given the physical conditions of UVABS, it would be detected as a
typical UV absorber if seen projected against the continuum source and BLR. 
We note that the observed UV spectrum of NGC~4151 shows components of
absorption that span a wide range in ionization and column density (Kraemer et
al. 2001b), but not a component that specifically matches our UVABS model
component for the central emission-line knot. This is likely due to the fact
that we are viewing the AGN at a special angle, close to the edge of the
emission-line bicone (Crenshaw et al. 2000), which may explain its complex
absorption spectrum (Kraemer et al. 2001b). The UVABS component could
be ``hidden'' in the complex of absorption, specifically in the broad component
``D+E'', or just not present in the line of sight due to a variation in
absorption properties with polar angle (with respect to the accretion and/or
torus axes). Nevertheless, an important follow-up study would be to model the
emission-line spectra of central knots in other Seyfert galaxies, and compare
these with their absorption-line properties to test the generality of our
results.

A number of dynamical models invoke accretion-disk winds to explain the
intrinsic absorption (and broad-line emission) in AGN, using radiation pressure
(Murray et al. 1985; Proga et al. 2000) and/or centrifugal acceleration along
magnetic field lines (Blandford \& Payne 1982; Bottorff et al. 2000).
The absorbers in these models are located close to the BLR, which is only
light-days from the central continuum source (presumably the accretion disk
around the central supermassive black hole).
However, our results suggest that a large fraction,
and perhaps a majority, of UV absorbers are located in the inner NLRs of
Seyfert galaxies, at distances of tens of parsecs from their central continuum
sources, and are therefore not directly associated with accretion-disk winds.
This does not rule out an origin in the accretion disk, but other origins for
the outflowing gas, such as the torus (Krolik \& Kriss 1995) are also possible.
Once again, NGC~4151 appears to be an exceptional case, presumably as a result
of our special viewing angle. The broad component ``D+E''  is at a distance of
only $\sim$0.03 pc from the continuum source (Kraemer et al. 2001b), and may
be the best candidate for an accretion-disk wind in a Seyfert galaxy. There are
also outflowing absorbers at large distances from the nucleus ($\sim$700 and
$\sim$2100 pc) in NGC~4151, which may be associated with emission-line clouds
further out in the NLR or in the extended NLR (ENLR).
Some of the absorbers in other Seyfert galaxies may also arise in these more
distant regions.

Finally, we note that X-ray absorbers, which tend to have higher column
densities than UV absorbers and likely dominate the mass outflow rates, may
also be associated with the NLR. Spectra at the highest possible resolutions
with {\it HST}/STIS and the {\it Chandra X-ray Observatory} ({\it CXO})
indicate that the UV and X-ray absorption in individual Seyfert galaxies cover
very similar ranges in radial velocity (Crenshaw et al. 2003a, and references
therein). This suggests that the UV and X-ray absorbers may be spatially
colocated; for example the UV absorbers could be high-density knots in an X-ray
wind (Krolik \& Kriss 1995, 2001). Also, {\it CXO} observations of NGC~4151 and
NGC~1068 (Ogle et al. 2000, 2003) show extended X-ray emission-line regions 
that are colocated with their NLRs, which would likely be seen as X-ray
absorbers if projected against the continuum source. 
Further identification of components in both emission and absorption should lead
to much tighter constraints on the physical conditions, geometry, and dynamics
of the outflowing gas.

\acknowledgments

Some of the data presented in this paper were obtained from the Multimission
Archive at the Space Telescope Science Institute (MAST). STScI is operated by
the Association of Universities for Research in Astronomy, Inc., under NASA
contract NAS5-26555. This research has made use of NASA's Astrophysics Data
System.

\clearpage

\figcaption[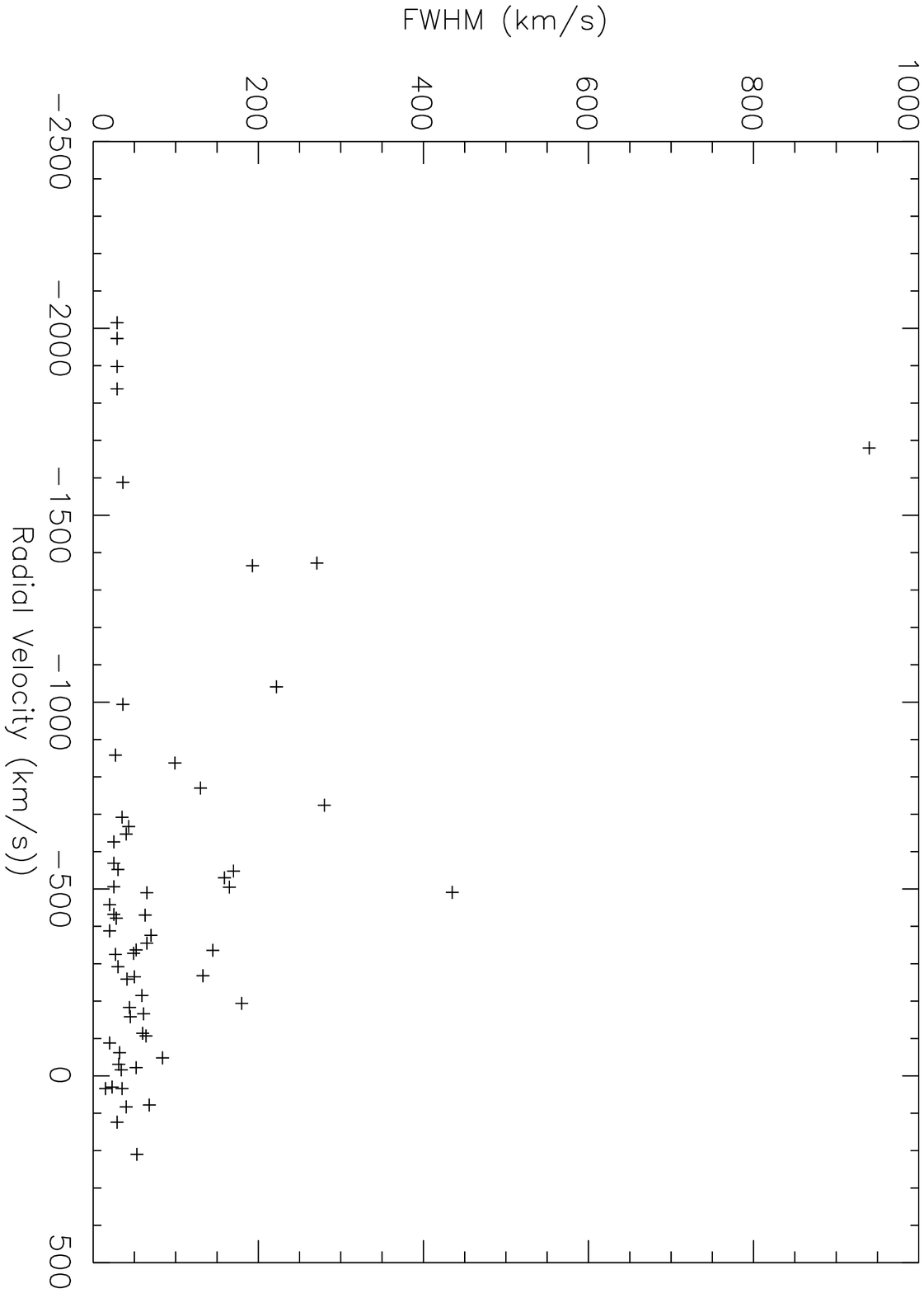]{Full-width at half-maximum (FWHM) vs. radial-velocity
centroid for individual components of intrinsic absorption in Seyfert 1
galaxies.}

\figcaption[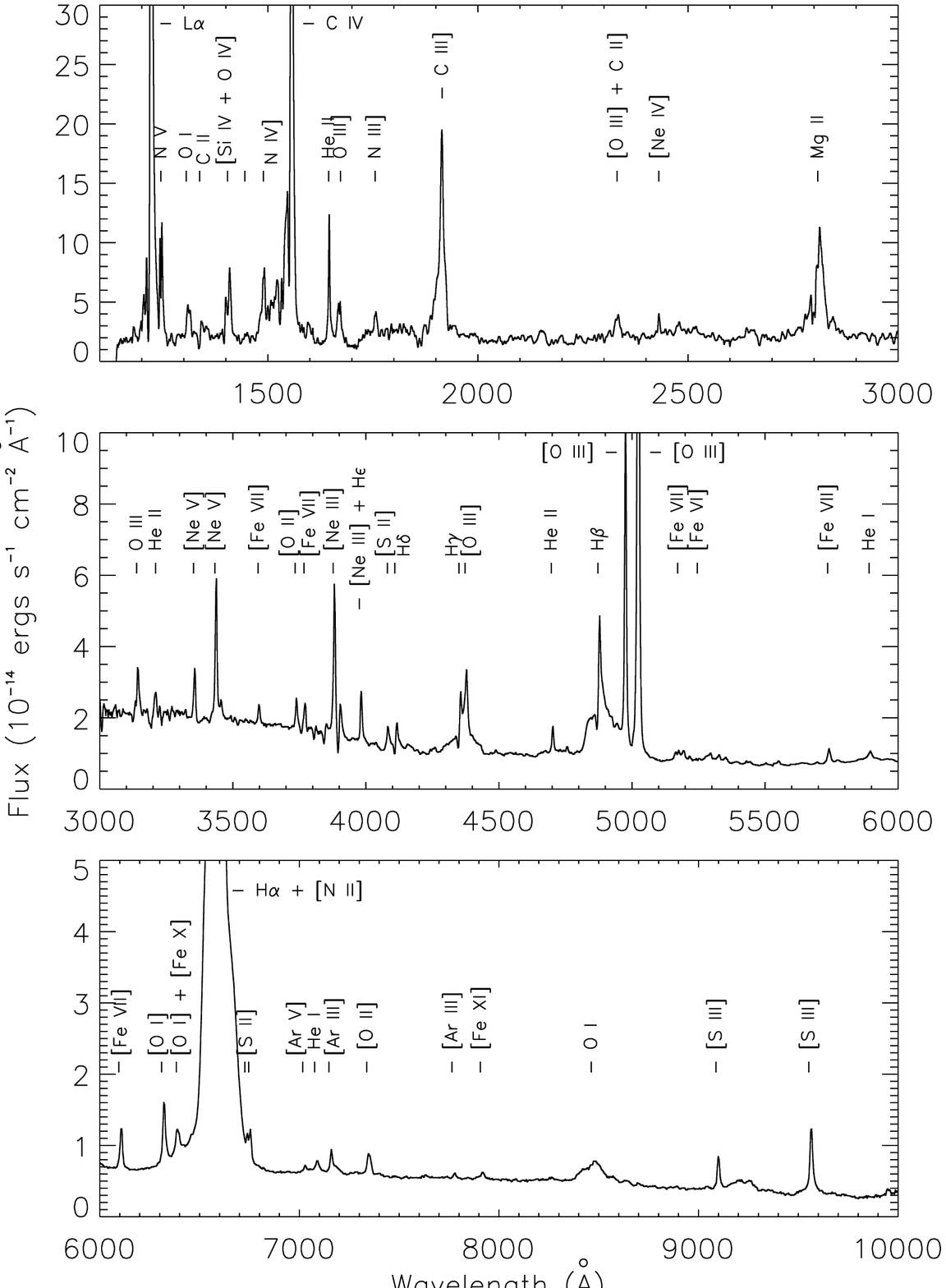]{STIS UV/optical spectrum of the central emission-line knot
in NGC~4151, obtained when the continuum and broad-line emission were in a low
state. Narrow emission lines that we detected and identified are labeled.}

\figcaption[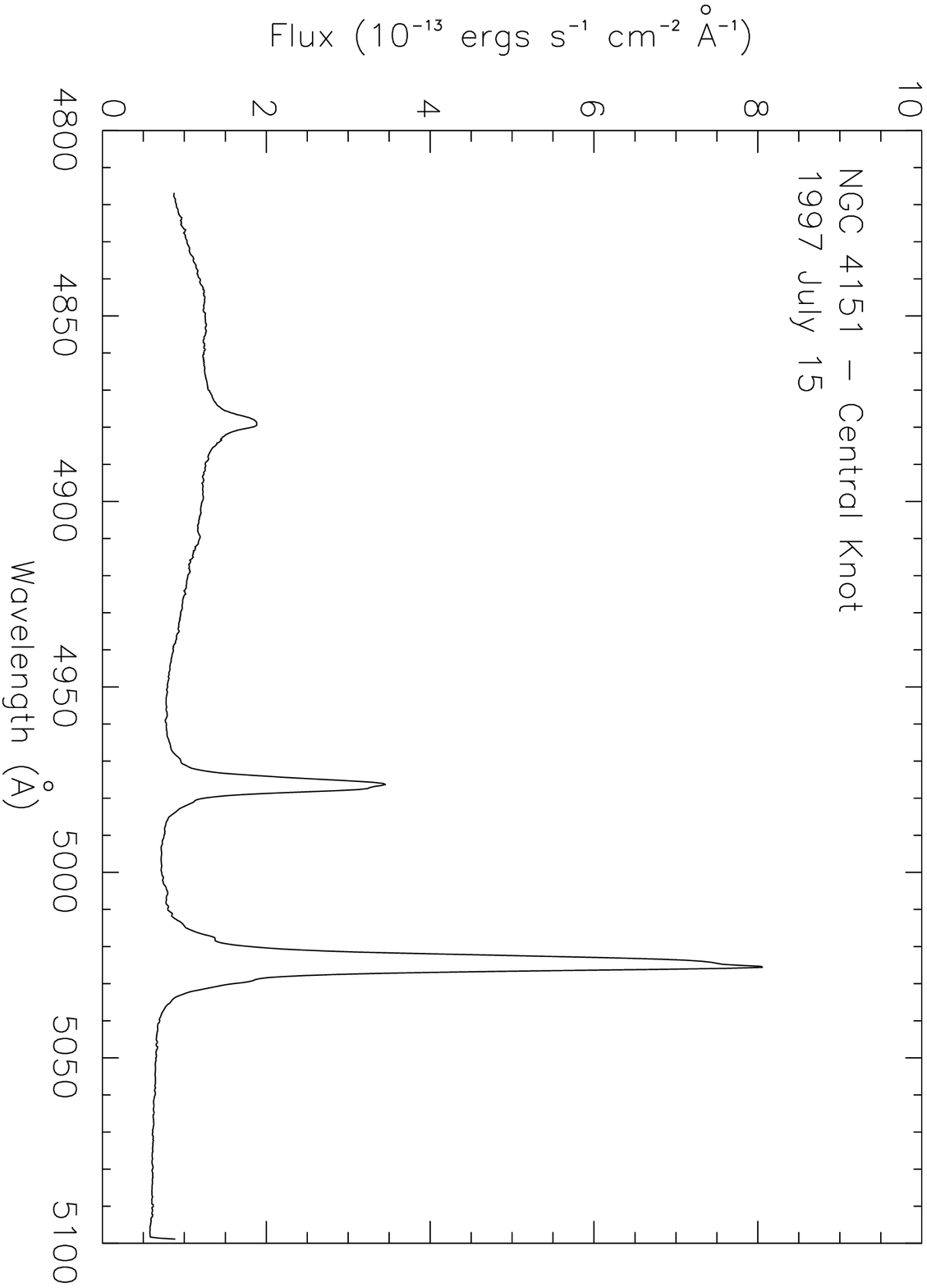]{Medium-dispersion high-state spectrum of the central
emission-line knot in NGC~4151, in the region around H$\beta$ and [O III]
$\lambda\lambda$4959, 5007.}

\clearpage

\begin{deluxetable}{lccccccc}
\tablecolumns{8}
\footnotesize
\tablecaption{Seyfert Galaxies with STIS Slitless Spectra \label{tbl-1}}
\tablewidth{0pt}
\tablehead{\colhead{Object} & \colhead{Seyfert} & \colhead{Redshift} & 
\colhead{Spatial} & \colhead{Spatial} & \colhead{Velocity} &
\colhead{ Velocity} \\  
\colhead{} & \colhead{Type} & \colhead{} &
\colhead{HWHM} & \colhead{HWHM} & \colhead{HWHM} & 
\colhead{HWZI} \\
\colhead{} & \colhead{} & \colhead{} &
\colhead{($''$)} & \colhead{(pc)$^a$} & \colhead{(km s$^{-1}$)} & 
\colhead{(km s$^{-1}$)}
}
\startdata
Mrk 573   &  2   & 0.0172 & 0.20 &  66 & 100 &  340 \\ 
Mrk 620   &  2   & 0.0061 & 0.14 &  17 &  90 &  310 \\
NGC 1386  &  2   & 0.0031 & 0.24 &  14 & 440 &  860 \\
NGC 4151  & 1    & 0.0033 & 0.11 &   7 & 150 & 1060 \\ 
NGC 3081  &  2   & 0.0079 & $\leq$0.07 &  $\leq$11 & 100 &  500 \\
NGC 3516  &  1   & 0.0083 & $\leq$0.07 &  $\leq$11 & 110 &  670 \\
NGC 5506  &  1.9 & 0.0061 & $\leq$0.07 &   $\leq$8 &  60 &  800 \\
NGC 5643  &  2   & 0.0040 & 0.16 &  12 & 110 &  680 \\ 
NGC 5728  &  2   & 0.0093 & 0.20 &  36 & 140 &  730 \\
NGC 7212  &  2   & 0.0267 & 0.13 &  67 & 260 & 1010 \\
\enddata
\tablenotetext{a}{For H$_0$ $=$ 75 km s$^{-1}$ Mpc$^{-1}$}
\end{deluxetable}

\begin{deluxetable}{ccrcccr}
\tablecolumns{7}
\scriptsize
\tablecaption{STIS Spectra of the Nuclear Knot in NGC~4151 \label{tbl-2}}
\tablewidth{0pt}
\tablehead{\colhead{Date} & \colhead{Detector} & \colhead{Grating}
& \colhead{Range} & \colhead{Aperture} & \colhead{P.A.$^{a}$} &
\colhead{Exposure}\\
\colhead{(UT)} & \colhead{} & \colhead{} & \colhead{(\AA)}
& \colhead{($''$)} & \colhead{(\deg)} & \colhead{(s)}}
\startdata
2000 May 28 & MAMA&  G140L &  1150 -- 1717  & 52 x 0.1 & 83 & 2220 \\
2000 May 24 & CCD & G230LB &  1672 -- 3077  & 52 x 0.1 & 83 & 2825 \\
2000 May 24 & CCD &  G430L &  2906 -- 5715  & 52 x 0.1 & 83 & 780 \\
2000 May 24 & CCD &  G750L &  5277 -- 10269 & 52 x 0.1 & 83 & 780 \\ 
1997 July 15 & CCD & G430M &  4817 -- 5098  & 52 x 52  & 45 & 2139 \\
1997 July 15 & CCD & G750M &  6294 -- 6862  & 52 x 52  & 45 & 1860 \\
\enddata
\tablenotetext{a}{Position angle of the cross-dispersion direction.}
\end{deluxetable}

\clearpage

\begin{deluxetable}{lcccccc}
\tablecolumns{7}
\footnotesize
\tablecaption{NGC 4151 Nuclear Knot-- Narrow Emission-Line Ratios (Relative to
H$\beta$$^a$) 
\label{tbl-3}}
\tablewidth{0pt}
\tablehead{
\multicolumn{1}{c}{}    & \multicolumn{2}{c}{Observations} &
 \multicolumn{4}{c}{Models} \\
\colhead{Emission Line} & \colhead{Observed}& \colhead{Dereddened$^b$} &
\colhead{Composite$^c$} & \colhead{UVABS$^{d,e}$}  &  \colhead{OIII$^f$} &
\colhead{DENSE$^g$}
}
\startdata
N IV] $\lambda$1486  &0.75$\pm$0.14 &0.81$\pm$0.20 &0.67 &1.82 &0.32 &0.65 \\
He II $\lambda$1640 &2.00$\pm$0.32 &2.15$\pm$0.46 &2.22 &6.88 &1.58 &1.60 \\
O III] $\lambda$1663 &0.50$\pm$0.08 &0.54$\pm$0.12 &1.05 &0.17 &0.57 &1.63 \\
N III] $\lambda$1750 &0.44$\pm$0.07 &0.47$\pm$0.10 &0.45 &0.15 &0.26 &0.67 \\
C III] $\lambda$1909 &3.12$\pm$0.46 &3.39$\pm$0.72 &4.77 &1.27 &3.08 &6.90 \\
O III]/C II] $\lambda$2324 &0.38$\pm$0.08 &0.41$\pm$0.11 &0.58 &--- &0.34 
&0.90\\
$[$Ne IV] $\lambda$2423 &0.44$\pm$0.08 &0.47$\pm$0.10 &0.35 &1.34 &0.39 &0.09 \\
O III $\lambda$3133 &0.70$\pm$0.11 &0.72$\pm$0.12 & & & & \\
He II $\lambda$3204 &0.18$\pm$0.05 &0.19$\pm$0.05 &0.12 &0.35 &0.08 &0.09 \\
$[$Ne V] $\lambda$3346 &0.55$\pm$0.10 &0.56$\pm$0.10 &0.38 &1.94 &0.15 &0.19 \\
$[$Ne V] $\lambda$3424 &1.66$\pm$0.23 &1.69$\pm$0.24 &1.03 &5.23 &0.41 &0.50 \\
$[$Fe VII] $\lambda$3588 &0.21$\pm$0.04 &0.21$\pm$0.04 & & & & \\
$[$O II] $\lambda$3727 &0.33$\pm$0.09 &0.34$\pm$0.09 &0.11 &--- &0.25 &0.02 \\
$[$Fe VII] $\lambda$3760 &0.33$\pm$0.07 &0.33$\pm$0.07 & & & & \\
$[$Ne III] $\lambda$3869 &1.92$\pm$0.27 &1.94$\pm$0.27 &1.92 &0.07 &1.69 &2.54 
\\
$[$Ne III] $\lambda$3967, H$\epsilon$ &0.56$\pm$0.09 &0.56$\pm$0.09 & & & & \\
$[$S II] $\lambda$4072 &0.33$\pm$0.08 &0.33$\pm$0.08 &0.26 &--- &0.13 &0.32 \\
H$\delta$ $\lambda$4100 &0.43$\pm$0.09 &0.44$\pm$0.09  &0.26 &0.26 &0.26 &0.26
\\
H$\gamma$ $\lambda$4340 &0.75$\pm$0.15 &0.75$\pm$0.16 &0.47 &0.47 &0.47 &0.47 \\
$[$O III] $\lambda$4363 &0.87$\pm$0.16 &0.88$\pm$0.16 &0.92 &0.05 &0.35 &1.56 \\
He II $\lambda$4686 &0.30$\pm$0.05 &0.30$\pm$0.05 &0.30 &0.89 &0.22 &0.21 \\
H$\beta$ $\lambda$4861 &1.00 &1.00  &1.00 &1.00 &1.00 &1.00 \\
$[$O III] $\lambda$4959 &3.64$\pm$0.46 &3.63$\pm$0.46 &4.53 &0.46 &7.21 &3.48 \\
$[$O III] $\lambda$5007 &12.17$\pm$1.55 &12.14$\pm$1.55 &13.61 &1.39 &21.65 
&10.43 \\
$[$Fe VII] $\lambda$5159 &0.12$\pm$0.02 &0.12$\pm$0.02 & & & & \\
$[$Fe VI] $\lambda$5176 &0.14$\pm$0.02 &0.14$\pm$0.02 & & & & \\
$[$Fe VII] $\lambda$5721 &0.24$\pm$0.04 &0.24$\pm$0.04 & & & & \\
He I $\lambda$5876 &0.09$\pm$0.03 &0.09$\pm$0.03 &0.11 &--- &0.11 &0.13 \\
$[$Fe VII] $\lambda$6087 &0.42$\pm$0.06 &0.41$\pm$0.07 &0.15 &0.29 &0.12 &0.14 
\\
$[$O I] $\lambda$6300 &0.63$\pm$0.10 &0.62$\pm$0.11 &0.55 &--- &0.39 &0.80 \\
$[$O I], [Fe X] $\lambda$6374 &0.33$\pm$0.06 &0.32$\pm$0.06 &0.25 &0.50 &0.13 
&0.27 \\
$[$N II] $\lambda$6548 &0.50$\pm$0.14 &0.49$\pm$0.14 &0.25 &--- &0.37 &0.23 \\
H$\alpha$ $\lambda$6563 &3.50$\pm$0.66 &3.42$\pm$0.67 &2.91 &2.71 &2.89 &2.98\\
$[$N II] $\lambda$6584 &1.50$\pm$0.42 &1.47$\pm$0.42 &0.76 &--- &1.11 &0.68 \\
$[$S II] $\lambda$6716 &0.08$\pm$0.01 &0.07$\pm$0.01 &0.04 &--- &0.09 &0.02 \\
$[$S II] $\lambda$6731 &0.18$\pm$0.02 &0.18$\pm$0.02 &0.10 &--- &0.19 &0.05 \\
$[$Ar V] $\lambda$7005 &0.07$\pm$0.01 &0.06$\pm$0.01 &0.03 &0.01 &0.03 &0.04 \\
He I $\lambda$7065 &0.10$\pm$0.01 &0.09$\pm$0.01 &0.08 &--- &0.08 &0.09 \\
$[$Ar III] $\lambda$7136 &0.24$\pm$0.03 &0.23$\pm$0.04 &0.23 &--- &0.25 &0.26 \\
$[$ O II] $\lambda$7325 &0.29$\pm$0.04 &0.28$\pm$0.05 &0.22 &--- &0.20 &0.28 \\
$[$ Ar III] $\lambda$7751 &0.06$\pm$0.01 &0.06$\pm$0.01 &0.05 &--- &0.06 &0.06
\\
$[$Fe XI] $\lambda$7892 &0.10 $\pm$0.02 &0.10$\pm$0.03 &0.01 &0.08 &--- &--- \\
$[$S III] $\lambda$9069 &0.32$\pm$0.05 &0.31$\pm$0.05 &0.35 &--- &0.55 &0.29 \\
$[$S III] $\lambda$9532 &0.91$\pm$0.15 &0.88$\pm$0.16 &0.88 &0.01 &1.37 &0.71 \\
\enddata
\tablenotetext{a}{Observed flux of H$\beta$ from entire knot (G430M spectrum)
$=$ 3.0 ($\pm$0.4) $\times$ 10$^{-13}$ ergs s$^{-1}$ cm$^{-2}$}
\tablenotetext{b}{Used E$_{B-V}$ $=$ 0.02 and standard Galactic reddening curve
(Savage \& Mathis 1979.}
\tablenotetext{c}{Weighted according to the fraction of H$\beta$ from each model
component: 0.12 (UVABS), 0.38 (OIII), 0.50 (DENSE).}
\tablenotetext{d}{$U =$ 0.03, $n_H$ $=$ 10$^{4.3}$ cm$^{-3}$,  $N_H$ $=$
10$^{19.5}$ cm$^{-2}$, d $=$ 9.5 pc, f$_{H\beta}$ $=$ 5.6 $\times$ 10$^{-2}$
ergs s$^{-1}$ cm$^{-2}$, $C_g$ $=$ 1.1.}
\tablenotetext{e}{Ionic column densities for UVABS are N(H~I) $=$ 4.5 $\times$
10$^{15}$ cm $^{-2}$, N(C~IV) $=$ 1.5 $\times$ 10$^{15}$ cm $^{-2}$, N(N~V) $=$
1.3 $\times$ 10$^{15}$ cm $^{-2}$, N(O~VI) $=$ 5.3 $\times$ 10$^{15}$ cm
$^{-2}$, and N(Si~IV) $=$ 3.1 $\times$ 10$^{12}$ cm $^{-2}$.}
\tablenotetext{f}{$U =$ 0.01, $n_H$ $=$ 10$^{4.8}$ cm$^{-3}$,  radiation
bound, d $=$ 9.5 pc, f$_{H\beta}$ $=$ 8.1 ergs s$^{-1}$ cm$^{-2}$, $C_g$ $=$
0.02}
\tablenotetext{g}{$U =$ 0.01, $n_H$ $=$ 10$^{6.2}$ cm$^{-3}$,  radiation
bound, d $=$ 2.0 pc, f$_{H\beta}$ $=$ 2.1 $\times$ 10$^{2}$ ergs s$^{-1}$
cm$^{-2}$, $C_g$ $=$ 0.03}
\end{deluxetable}


\clearpage
\begin{figure}
\plotone{f1.eps}
\\Fig.~1.
\end{figure}

\clearpage
\begin{figure}
\epsscale{0.85}
\plotone{f2.eps}
\epsscale{1.0}
\\Fig.~2.
\end{figure}

\clearpage
\begin{figure}
\plotone{f3.eps}
\\Fig.~3.
\end{figure}


\clearpage
\begin{references}

\reference{bro1985}Bromage, G.E., et al. 1985, \mnras, 215, 1

\reference{cec2002}Cecil, C., Dopita, M.A., Groves, B., Wilson, A.S., Ferruit,
P., P\'{e}contal, E., \& Binette, L. 2002, \apj, 568, 627

\reference{col2001}Collinge, M.J., et al. 2001, \apj, 557, 2

\reference{cre1999}Crenshaw, D.M., Kraemer, S.B., Boggess, A., Maran, S.P.,
Mushotzky, R.F., \& Wu, C.-C. 1999, \apj, 516, 750

\reference{cre2004a}Crenshaw, D.M., Kraemer, S.B., \& Gabel, J.R. 2004a, in AGN
Physics with the Sloan Digital Sky Survey, ed. G.T. Richards \& P.B. Hall (San
Francisco: ASP), ASP Conference Series, 311, 235

\reference{cre2004b}Crenshaw, D.M., Kraemer, S.B., Gabel, J.R., Schmitt, H.R.,
Filippenko, A.V., Ho, L.C., Shields, J.C., \& Turner, T.J. 2004b, \apj, 612, 152

\reference{cre2003a}Crenshaw, D.M., Kraemer, S.B., \& George, I.M. 2003a, 
\araa, 41, 117

\reference{cre1986}Crenshaw D.M., \& Peterson, B.M. 1986, \pasp, 98, 185.

\reference{cre2000}Crenshaw, D.M., et al. 2000, \aj, 120, 1731

\reference{cre2002}Crenshaw, D.M., et al. 2002, \apj, 566, 187.

\reference{cre2003b}Crenshaw, D.M., et al. 2003b, \apj, 594, 116.

\reference{esp1998}Espey, B.R., Kriss, G.A., Krolik, J.H., Zheng, W., Tsvetanov,
Z., \& Davidsen, A.F. 1998, \apj, 500, L13

\reference{fer1998}Ferland, G.J., Korista, K.T., Verner, D.A., Ferguson, J.W.,
Kingdon, J.B., \& Verner, E.M. 1998, \pasp, 110, 761

\reference{gab2003a}Gabel, J.R., et al. 2003a, \apj, 583, 178.

\reference{gab2003b}Gabel, J.R., et al. 2003b, \apj, 595, 120

\reference{gab2004}Gabel, J.R., et al. 2004, \apj, submitted.

\reference{geo1998}George, I.M., Turner, T.J., Netzer, H., Nandra, K.,
Mushotzky, R.F., \& Yaqoob, T. 1998, \apjs, 114, 73

\reference{gre1989} Grevesse, N. \& Anders, E. 1989, in Cosmic Abundances of 
Matter, ed. C.J. Waddington (New York: AIP),1 

\reference{hut2002}Hutchings, J.B., Crenshaw, D.M., Kraemer, S.B., Gabel, J.R.,
Kaiser, M.E., Weistrop, D., \& Gull, T.R. 2002, \aj, 124, 2543.

\reference{hut1998}Hutchings, J.B., et al. 1998, \apj, 492, L115

\reference{kai2000}Kaiser, M.E., et al. 2000, \apj, 528, 260 

\reference{kra2000a}Kraemer, S.B. \& Crenshaw, D.M. 2000a, ApJ, 532, 256.

\reference{kra2000b}Kraemer, S.B. \& Crenshaw, D.M. 2000b, \apj, 544, 763

\reference{kra2001a}Kraemer, S.B., Crenshaw, D.M., \& Gabel, J.R. 2001a, \apj,
557, 30.

\reference{kra2002}Kraemer, S.B., Crenshaw, D.M., George, I.M., Netzer, H.,
Turner, T.J., \& Gabel, J.R. 2002, \apj, 577, 98.

\reference{kra2001b}Kraemer, S.B., Crenshaw, D.M., Hutchings, J.B., 
Danks, A.C., Gull, T.R., Kaiser, M.E., Nelson, C.H., \& Weistrop, D. 2001b, 
\apj, 551, 671.

\reference{kra2000b}Kraemer, S.B., Crenshaw, D.M., Hutchings, J.B., Gull, T.R.,
Kaiser, M.E., Nelson, C.H., \& Weistrop, D. 2000, \apj, 531, 278.

\reference{kra2003}Kraemer, S.B., Crenshaw, D.M., Yaqoob, T., McKernan, B.,
Gable, J.R., George, I.M., Turner, \& Dunn, J.P. 2003, \apj, 582, 125.

\reference{kri2002}Kriss, G.A. 2002, in Mass Outflow in Active Galactic Nuclei:
New Perspectives, ed. D.M. Crenshaw, S.B. Kraemer, \& I.M. George (San
Francisco: ASP), ASP Conference Series, 255, 69

\reference{kri2003}Kriss, G.A., Blustin, A., Branduardi-Raymont, G., Green,
R.F., Hutchings, J., \& Kaiser, M.E. 2003, A\&A, 403, 473

\reference{kri1997}Kriss, G.A., Krolik, J., Grimes, J., Tzvetanov, Z., Espey,
B., Zheng, W., \& Davidsen, A. 1997, in Emission Lines in Active Galaxies: New
Methods and Techniques, ed. B.M. Peterson, F.-Z. Cheng, \& A.S. Wilson, (San
Francisco: ASP), ASP Conference Series, 113, 453

\reference{kri2000}Kriss, G.A., et al. 2000, \apj, 538, L17

\reference{kro1995}Krolik, J.H. \& Kriss, G.A. 1995, \apj, 447, 512

\reference{kro2001}Krolik, J.H. \& Kriss, G.A. 2001, \apj, 561, 684

\reference{mat1995}Mathur, S., Elvis, M., \& Wilkes, B. 1995, \apj, 452, 230

\reference{nel2000}Nelson, C.H., Weistrop, D., Hutchings, J.B., Crenshaw, D.M.,
Gull, T.R., Kaiser, M.E., Kraemer, S.B., \& Lindler, D. 2000, \apj, 531, 257.

\reference{net2002}Netzer, H., Chelouche, D., George, I.M., Turner, T.J.,
Crenshaw, D.M., Kraemer, S.B., \& Nandra, K. 2002, \apj, 571, 256

\reference{net1993}Netzer, H. \& Laor, A. 1993, \apj, 404, L51

\reference{net2003}Netzer, H., et al. 2003, \apj, 599, 933.

\reference{ogl2000} Ogle, P.M., et al. 2000, A\&A, 402, 849

\reference{ogl2000} Ogle, P.M., et al. 2003, \apj, 545, L81

\reference{oli1997}Oliva, E. 1997, in ASP Conference Ser. 113, Emission Lines in
Active Galaxies: New Methods and Techniques, ed. B.M Peterson, F.-Z. Cheng, \&
A.S. Wilson (San Francisco: ASP), 288

\reference{ost1989}Osterbrock, D.E. 1989, Astrophysics of Gaseous Nebulae and
Active Galactic Nuclei (University Science Books: Mill Valley)

\reference{rey1997}Reynolds, C.S. 1997, \mnras, 286, 513

\reference{rui2004}Ruiz, J.R., Crenshaw, D.M., Kraemer, S.B., Bower, G.A., 
Gull, T.R., Hutchings, J.B., Kaiser, M.E. and Weistrop, D. 2005, \aj, in press
(astro-ph/0409754)

\reference{sav1979}Savage, B.D., \& Mathis, J.S. 1979, ARAA, 17, 73

\reference{sch1998}Schlegel, D.J., Finkbeiner, D.P., \& Davis, M. 1998, \apj, 
500, 525

\reference{sch2003a}Schmitt, H.R., Donley, J.L., Antonucci, R.R.J., Hutchings, 
J.B., \& Kinney, A.L. 2003a, \apjs, 148, 327

\reference{sch2003a}Schmitt, H.R., Donley, J.L., Antonucci, R.R.J., Hutchings, 
J.B., Kinney, A.L., \& Pringle, J.E. 2003b, \apj, 597, 768 

\reference{sco1994}Scott, J.E., et al. 2004, \apjs, 152, 1

\reference{wey1997}Weymann, R.J., Morris, S.K., Gray, M.E., \& Hutchings, J.B.
1997, \apj, 483, 717

\end{references}
\end{document}